\documentclass[a4paper,12pt,titlepage,twoside]{article}

\usepackage{graphicx}        
\usepackage{graphics}
\usepackage{color}
\usepackage{a4}               
\usepackage{setspace}

\title{An approximate model for the adhesive contact of rough viscoelastic surfaces}
\author{Guillaume Haiat$^1$ and Etienne Barthel$^2$\\\\ 1\_ CNRS UMR 7052 - B2OA Laboratoire
de M\'ecanique Physique\\61, Avenue du G\'en\'eral de Gaulle 94010
Cr\'eteil - France\\2\_ Laboratoire CNRS/Saint-Gobain "Surface du
Verre et Interfaces",\\39, quai Lucien Lefranc, BP 135, F-93303
Aubervilliers Cedex,France.}

\newcommand{\dsd}[1]{\frac{\partial}{\partial #1}}
\newcommand{\dsdd}[1]{\frac{d}{d #1}}
\newcommand{\tam}[1]{t_{a-}(#1)}
\def\psd{\frac{\pi}{2}}
\def\pst{\frac{\pi}{3}}

\def\psit{\tilde\psi}
\def\phit{\tilde\phi}

\begin{document}

\maketitle

{\LARGE{Abstract}}

Surface roughness is known to easily suppress the adhesion of
elastic surfaces. Here a simple model for the contact of
\emph{viscoelastic} rough surfaces with significant levels of
adhesion is presented. This approach is derived from our previous
model [E.~Barthel and G.~Haiat {\em Langmuir}, 18 9362 2002] for the
adhesive contact of viscoelastic spheres. For simplicity a simple
loading/unloading history (infinitely fast loading and constant
pull-out velocity) is assumed. The model provides approximate
analytical expressions for the asperity response and exhibits the
full viscoelastic adhesive contact phenomenology such as stress
relaxation inside the contact zone and creep at the contact edges.
Combining this model with a Greenwood-Williamson statistical
modeling of rough surfaces, we propose a quantitative assessment of
the adhesion to rough viscoelastic surfaces. We show that moderate
viscoelasticity efficiently restores adhesion on rough surfaces over
a wide dynamic range.
\newpage

\section{Introduction}
{Surface roughness is a major parameter for the control of the
adhesion of viscoelastic materials and as such is central to many
technological processes. As an example, in the automotive industry,
when assembling a polymer interlayer with two glass sheets to form a
laminated windshield, a periodic roughness is intentionally
generated on the polymer to reduce the adhesion and facilitate
positioning before the final heat treatment~\cite{JUANG01}.
Roughness also plays a role in the self-adhesion mechanism (or tack)
of soft polymers~\cite{ZOSEL97,GAY99} like repositionable adhesives.
In a similar way, in the glass industry, glass molding is performed
at elevated temperatures for easy glass flow but at the risk of
adhesion. Indeed adhesion of hot glass to the mold halts production
until the mold has been cleaned of its adherents. The roughness of
the mold strongly impacts the adhesion of the viscoelastic glass
gob~\cite{PECH05} and a stochastic but carefully tailored roughness,
obtained through shot peening of the steel surfaces for instance,
will promote an easy molding process.}

The adhesive contact to rough viscoelastic surfaces is a complex
problem however because it couples the statistical approach of the
adhesive contact to \emph{rough surfaces} with the difficulties
involved in the \emph{viscoelastic} adhesive contact. For
\emph{elastic} solids, the adhesive contact to rough surfaces is
reasonably well understood. Building upon the Greenwood-Williamson
independent asperity approach, Fuller and Tabor~\cite{Fuller75} have
given a useful description of the strong reduction of the adhesive
force incurred when roughness increases. They introduce an adhesion
parameter $\sigma_s/\delta_c$ where $\sigma_s$ is the standard
deviation of the {summit height distribution} and $\delta_c$ is the
maximum single asperity extension before rupture. This adhesion
parameter is a measure of the energy balance between stretched and
compressed asperities. If the adhesion parameter is large, the
compression of the upper asperities dominates the traction of the
adhesive lower asperities which significantly depresses the
adhesion.

The calculations are relatively straightforward for elastic contacts
which are essentially reversible. For viscoelastic contacts,
adhesion depends upon the full contact history so that the
calculations are more involved. In the absence of adhesion, the
viscoelastic contact problem was first solved by Ting~\cite{Ting66}.
Subsequently, important results for the viscoelastic \emph{crack}
problem were obtained by Schapery~\cite{Schapery75,Schapery75bis}
and Greenwood and Johnson~\cite{Greenwood81,Greenwood04}. Hui and
coworkers~\cite{Hui98,Lin99,Hui00} have tackled the problem of
viscoelastic \emph{adhesive} contacts. We recently proposed a model
for the full viscoelastic contact of adhesive
bodies~\cite{Barthel02,Haiat03,Barthel04}.

The present paper inquires into the application of this model to the
adhesive contact to rough surfaces. To model the response of a
distribution of asperities a simplified description of the
viscoelastic contact is needed and the first part of the paper
develops such a minimal model. It is devised so that the two main
phenomena characteristic for viscoelastic adhesive contact -- stress
relaxation under the contact and creep induced by the adhesive
interactions at the contact edges -- are preserved while keeping
calculations to a minimum. The model exhibits typical viscoelastic
adhesive contact phenomenology, and in particular the time lag
between indenter retraction and the actual contact radius recession,
also called "stick period"~\cite{Barthel02}. In the second part of
the paper {examples} of the use of this approximate model for rough
surfaces are presented. The impact of the pull-out velocity on the
pull-out force for typical rough surfaces is calculated. The results
emphasize how a moderately viscoelastic material effectively
restores the adhesion lost through roughness.
\section{Model} \label{Sec_Model}
\subsection{Single asperity contact} \label{Sec_Single_Asperity_Contact}
Modeling the viscoelastic contact of an {axisymmetric} asperity in
the presence of adhesion is complex because, as in all viscoelastic
problems, the solution depends upon the full contact history.

In addition, for contact problems, one must deal with mixed boundary
conditions: the surface normal displacement $u(r)$ is prescribed
inside the contact zone whereas the normal surface stress
$\sigma(r)$ due to the adhesive interactions is specified inside the
so-called cohesive zone (Figure~\ref{Fig_Contact_Model}), outside
the contact zone. Moreover the boundary between these two regions is
not fixed but moves during the contact history. For a viscoelastic
material, the solution is reasonably simple for growing contact
radius~\cite{Hui98,Schapery89}. However, during pull-out ({\it i.e.}
decreasing contact area), the residual deformation has to be taken
into account. This in turn depends upon the full stress history,
which leads to a complex situation, both for
adhesionless~\cite{Ting66} and
adhesive~\cite{Hui98,Lin99,Hui00,Barthel02,Haiat03} contacts.

\subsection{Resolution method}\label{Sec_Resolution_Method}
Under the assumption of axial symmetry, we resort to two auxiliary
functions $g(r)$ and $\theta(r)$, which are the following Abel
transforms of respectively the  distributions of normal surface
stress $\sigma(r)$ and normal surface displacement $u(r)$:
\begin{eqnarray}
g(r) & = & - \int_{r}^{+\infty} \frac{s \sigma(s)}{\sqrt{s^2-r^2}}ds,  \label{defg} \\
\theta(r) & = & \dsdd{r}\int_0^r\frac{su(s)}{\sqrt{r^2-s^2}}ds,
\label{defth1}
\end{eqnarray}

\subsubsection{Boundary conditions}\label{Sec_Boundary_Conditions} A first benefit of these transforms
is their pertinence for {axisymmetric} boundary conditions. In
particular, inside the contact zone ($r<a$), the normal surface
displacement is known through the contact condition
\begin{equation}
  u(r)= \delta - h(r)
\end{equation}
where $\delta$ is the penetration and $h(r)$ the shape of the
indenter (Figure~\ref{Fig_Contact_Model}).
Then, using Eq.~\ref{defth1}, inside the contact zone
\begin{equation}\label{defth2}
\theta(r)  =   \delta - \delta_0(r)
\end{equation}
where
\begin{equation}
\delta_0(r) = \dsdd{r}\int_0^rds\frac{sh(s)}{\sqrt{r^2-s^2}}\cdot
\end{equation}
It turns out that $\delta_0(r)$ is the Hertz penetration for a
contact radius $r$ as shown in the next section. Note that this
function $\delta_0(r)$ is determined by the shape of the indenter
only. It is equal to $r^2/R$ for a sphere of radius $R$ and to $\psd
r/\tan\beta$ for a cone of apical angle $\beta$.

Conversely, as {mentioned} above, the adhesive stress distribution
$\sigma(r)$ outside the contact zone is known. More precisely, in
the present case, we will show in section~\ref{Sec_Cohesive_Zone}
that it can be self-consistently determined through an independent
set of equations, under the assumption of small cohesive zone size.
Therefore the $g$ function is known outside the contact zone
($r>a$).

\subsubsection{Mechanical Equilibrium -- Viscoelastic
materials}\label{Sec_Mechanical_Equilibrium} The second benefit of
these auxiliary functions is that they naturally handle the long
range nature of the elastic field: for instance, under the
conditions of linear elastic behaviour (Young's modulus $E$, Poisson
ratio $\nu$) and absence of shear stresses at the contact,
mechanical equilibrium leads to {\begin{equation}
 g(r)= \frac{E^\star}{2} \theta(r) \label{equel}
\end{equation}
where
\begin{equation}
E^\star = \frac{E}{(1-\nu^2)}
\end{equation}
}The contact problem is solved under the assumption of continuity of
the stress distribution at $a$. The penetration is then directly
obtained by~\cite{Barthel02} {\begin{equation}
  g(a)=\frac{E^\star}{2} \theta(a)
\end{equation}}
For {example} for an adhesionless contact, $\sigma(r)=0$ for $r\geq
a$ so that $g(a)=0$ and $\delta=\delta_0(a)$ as {mentioned} in
section~\ref{Sec_Boundary_Conditions}. If adhesion is present,
$g(a)$ is not zero and a more complex situation arises.

For a {\em linear viscoelastic} material, {under the approximation
of constant Poisson ratio~\cite{LAKES06}} we introduce the creep and
the relaxation functions
\begin{eqnarray}
\phi(t) & = & \frac{2}{E^*}\phit(t) \label{Eq_phi}\\
\psi(t) & = & \frac{E^*}{2} \psit(t) \label{Eq_psi}
\end{eqnarray}
with
\begin{eqnarray}
\phit(t) & = & 1+\frac{(1-k)}{k}(1-e^{\frac{-t}{\mu}})\label{Eq_phi_norm}\\
\psit(t) & = & k+(1-k) e^{\frac{-t}{k \mu}} \label{Eq_psi_norm}
\end{eqnarray}
where $\mu$ is the creep time. The parameter $k$ lies between $1$
(elastic) and $0$ (Maxwell). Due to the decoupling between spatial
and temporal responses, the equilibrium equation (Eq.~\ref{equel})
will now be
\begin{equation}
g(r,t)=\int_0^t d\tau \psi(t-\tau) \dsdd{\tau} \theta(r,\tau)
\label{gdth}
\end{equation}
or its inverse
\begin{equation}
 \theta(r,t)=\int_0^t d\tau \phi(t-\tau) \dsdd{\tau}
g(r,\tau) \label{thdg}
\end{equation}
where the relaxation function $\psi(t)$ and the creep function
$\phi(t)$ are inverse for this product of convolution.
 In the case of increasing contact radius, Eqs.~\ref{defth2} and
 \ref{thdg}
determine the penetration by
\begin{equation}
\delta(t)=\delta_0(a(t)) + \int_0^t d\tau\phi(t-\tau) \dsd{\tau}
g(a(t),\tau) \label{deplin}
\end{equation}
This result is equivalent to previous
formulations~\cite{Lin99,Schapery89}.

The decreasing contact radius case is more intricate. Taking into
account the domains where $g$ and $\theta$ are known respectively,
we have to resort to Eqs.~\ref{defth2}, \ref{gdth} and \ref{thdg} to
determine the penetration by the integral equation
\begin{equation}
g(a(t),t)=
\int_{\tam{a(t)}}^{t}d\tau\psi(t-\tau)\dsd{\tau}\left\{\delta(\tau)-\delta_0(a(t))\right\}+
\bar g(a(t),t)\label{deplout}
\end{equation}
Here $\tam{a(t)}$ is the time at which the contact radius was first
equal to $a(t)$, during the loading phase. The corrective term
\begin{equation}
  \bar g(a(t),t) = \int_0^{\tam{a(t)}} d\tau \psi(t-\tau)\dsd{\tau}\left( \int_0^\tau
d\tau' \phi(\tau-\tau')\dsd{\tau'}g(a(t),\tau') \right)
\end{equation}
is unessential, except for the cross-over between increasing and
decreasing contact radius regimes ($a=a_{max}$, $t=t_{max}$).

Then for a given history of the penetration, one can calculate the
history of the contact radius and subsequently the
force~\cite{Barthel02,Haiat03,Barthel04}. {Under the assumption of
small cohesive zone size (see section~\ref{Sec_Cohesive_Zone} for
the cohesive zone and section~\ref{Sec_Force_App} for details of the
calculation)} the force is given by Eq.~19 in \cite{Barthel02}:
\begin{equation}
P(t) =  4 \int_0^t d\tau \psi(t-\tau) \frac{d}{d\tau}
\int_0^{\min\left(a(t),a(\tau)\right)} dr
\left(\delta(\tau)-\delta_0(r)\right) \label{Eq_ForceRef}
\end{equation}
Note that this equation is valid for both inward and outward runs.
{Eqs.~\ref{deplout} and \ref{Eq_ForceRef} involve stress relaxation
inside the contact zone but also crack tip viscoelasticity through
$g(a(t),t)$, as discussed in more details in the next paragraph.}

\subsubsection{Cohesive Zone}\label{Sec_Cohesive_Zone}
{For any adhesive contact, in the context of the present (Sneddon)
method,} the quantity $g(a)$, which is the function $g$ evaluated at
the edge of the contact zone, plays a central role. For the contact
zone, it determines Eqs.~\ref{deplin}, \ref{deplout} and
\ref{Eq_ForceRef} between macroscopic contact variables. But we have
previously shown that, for the cohesive zone, $g(a)$ determines the
stress intensity factor~\cite{Barthel07} characteristic of the local
deformation process at the crack tip.

{Here we dicuss, for the specific case of a viscoelastic adhesive
contact, how $g(a(t),t)$ is determined by the cohesive zone
parameters. This results in an independent relation between $g(a)$
and the contact radius velocity $\dot a$ through
Eqs.~\ref{Eq_GAdhes}, \ref{Eq_SelfCon_Back}, \ref{Eq_phi1} and
\ref{Eq_dwell}.}

This problem is similar to a viscoelastic
crack~\cite{Schapery75,Schapery75bis,Greenwood81,Greenwood04} {and
calls for a consistent treatment of an adhesive-stress induced creep
inside a cohesive zone of finite size ($c(t)\geq r \geq a(t)$,
Figure~\ref{Fig_Contact_Model}).} The equivalence of various methods
for the description of the cohesive zone has been demonstrated
earlier~\cite{Barthel98}. In brief, the details of the description
of the stresses inside the cohesive zone is unimportant as long as
the effective range of these interactions is preserved. Greenwood
and Johnson~\cite{Greenwood98} have proposed a description (the
so-called "Double-Hertz") which is analytically simple and which we
have retained for our viscoelastic models~\cite{Barthel02}.
Following these authors, we assume a specific stress distribution
which results in
\begin{equation}
g(r)= \frac{\pi}{4} \sigma_0 \frac{r^2-c^2}{\sqrt{c^2-a^2}}.
\end{equation}

Assuming that the size of the cohesive zone is small compared to the
contact zone ($a\gg c-a\equiv\epsilon$) and taking the limit when $r
\rightarrow a$ gives:
\begin{equation}
g(a)= - \frac{\pi}{4} \sigma_0 \sqrt{2a\epsilon} \label{Eq_GAdhes}
\end{equation}
For a decreasing contact radius the self-consistent equation
coupling the total adhesion energy and the deformation of the
contact zone writes (Eq. 23 of Ref. \cite{Barthel02}):
\begin{equation}
w=\frac{\pi}{8}\sigma_0^2 \epsilon \phi_{1}(t_r)
\label{Eq_SelfCon_Back}
\end{equation}
where $w$ is the adhesion energy,
\begin{equation}\label{Eq_phi1}
\phi_{1}(t) = \frac{2}{t^2} \int_0^t d\tau \tau \phi(t-\tau)
\end{equation}
and $t_r$ is a fracture dwell time. For non vanishing contact radius
velocity, this is the time it takes for the crack tip to move a
distance equal to the cohesive zone size. It is given by:
\begin{equation}
t_r(t)=\frac{-\epsilon(t)}{\dot a(t)}   \label{Eq_dwell}
\end{equation}

The simultaneous resolution of Eqs.~\ref{Eq_GAdhes}-\ref{Eq_dwell}
determines the velocity dependence of $g(a)$. It is typical of
viscoelastic crack problems {: compare our set of
Eqs.~\ref{Eq_GAdhes}, \ref{Eq_SelfCon_Back} and \ref{Eq_dwell} with,
for instance, Eqs.~1, 47 and 48 in~\cite{Schapery75}. As a result,
$g(a)$ increases with $\dot a$ as displayed on
Figure~\ref{Fig_Outer_K}: again compare {\it e. g.} with Fig.~6~b
in~\cite{Greenwood04}.}

\subsubsection{General Model -- stick period}\label{Sec_Stick_Zone}
For the viscoelastic adhesive contact, the main phenomenon we have
identified \cite{Barthel02,Haiat03} is the "pinning" of the contact
upon unloading (Figure~\ref{Fig_Histoire_Pene}). More precisely, for
such viscoelastic contacts, there is a typical time lag between the
retraction of the indenter and the actual contact radius recession.
We have called "stick period" this period over which the contact
radius stays pinned, before contact edge retraction sets in, finally
leading to contact rupture.

The physical interpretation of the stick period phenomenon has
already been given in~\cite{Barthel02}. Relaxation of the stresses
{\em inside} the contact zone reduces the stress intensity factor
and hence $g(a)$. This results in a slow crack motion
(Figure~\ref{Fig_Outer_K}) and the contact radius practically comes
to a halt. When pulling the indenter out, enough tensile stresses
must be re-built inside the contact zone before $g(a)$ is large
enough for the contact edge to be restored into propagating
condition (Figure~\ref{Fig_Outer_K}), and this depends upon the
competition between stress relaxation and pull out velocity as
discussed in~\cite{Barthel04}.

\subsection{Approximate Model}
We need a simple description of the viscoelastic adhesive contact
for rough surface calculations. We restrict the developments to the
spherical asperity (radius $R$) usual in rough surface models
although extension to other asperity shapes should be
straightformard. We also introduce two major simplifications to the
full model. One is an assumption on the type of loading/unloading.
The other one is an approximation for the stick period.

\subsubsection{Simplifying assumption -- Direct unloading} \label{Descri} We consider
fixed-grip loading, {\it i.e.} a loading where the penetration is
prescribed. In addition, the loading phase is considered to be
infinitely fast and the pull-out velocity $\dot{\delta}$ a
(negative) constant (Figure~\ref{Fig_Histoire_Pene}), so that the
penetration history simply writes:
\begin{equation}\label{Eq_Depl_Unload}
\delta(t) = (\delta_0 + t\dot{\delta})\Theta(t)
\end{equation}
where $\delta_0$ is the initial penetration\footnote{ This initial
penetration cannot be confused with the $\delta_0(r)$ function
introduced earlier and which will now be substituted by its specific
value for a sphere $r^2/R$ in the rest of the paper.}, taken
positive when the two surfaces are pushed one into another. $\Theta$
is the Heaviside step function. {Due to the infinitely fast loading
phase the quantity $t_{a-}(a(t))$ defined in the unloading phase
(section~\ref{Sec_Mechanical_Equilibrium}) is equal to zero.}

\subsubsection{Stick period approximation}
This approximation stems from the observation that for large
penetrations, the contact radius upon loading is primarily
determined by the instantaneous elastic response and that crack tip
creep during loading is not a dominant process~\cite{Barthel02}.

As a result, in the following, the stick period is approximately
modeled by a period of constant contact radius prescribed by the
instantaneous modulus. This will be a good approximation except for
small initial contact radii where the adhesive term is dominant.

In summary the approximate model is as follows:
\begin{enumerate}
  \item calculate the initial contact radius from an {\it elastic} model
with the instantaneous Young's modulus; assume this contact radius
stays constant to the end of the stick period. The main viscoelastic
process at work here is the stress relaxation inside the contact
zone, which directly impacts the force
(section~\ref{Sec_Force_Appro});
  \item determine if there is a stick period and if so calculate ending time $t_f$ (section~\ref{Sec_End_Stick_Period});
  \item calculate the unloading phase (section~\ref{Sec_Unloading}) and in particular the adhesion force.
\end{enumerate}

\subsubsection{Stick Period -- Force}\label{Sec_Force_Appro}

The force in the stick period is obtained from Eq.~\ref{Eq_ForceRef}
taking into account the discontinuity of the penetration at
$\tau=0$. This leads to {(section~\ref{Sec_Force_App_Stick})}:
\begin{equation}\label{Eq_ForceStick}
P_1(t)= 4 \psi(t)\left(\frac{2a_0^3}{3R}- \frac{\sigma_0 \pi
a_0^{3/2}\sqrt{\epsilon_0}}{\sqrt{2}E^*} \right) +4 a_0\dot\delta
\int_0^t d\tau \psi(t-\tau)
\end{equation}
which can be rewritten as:
\begin{equation}
P_1(t)= \psit(t)P_0 +4 a_0\dot\delta \int_0^t d\tau \psi(t-\tau).
\end{equation}
where the normalized form of $\psi$ (Eq.~\ref{Eq_psi_norm}) has been
used. $P_0$ is the initial force{ and $\epsilon_0$ the cohesive zone
size at $t=0$ which can be calculated from
Eqs.~\ref{Eq_GAdhes}-\ref{Eq_SelfCon_Back}}. The first term results
from the relaxation of the initial force and the second one is due
to the flat punch viscoelastic response and is proportional to the
(constant) pull-out velocity.

\subsubsection{Approximate determination of the end of the stick period} \label{Sec_End_Stick_Period}
{At the end of the stick period $a(t_f)=a_0$ because the contact
radius increase has been neglected by assumption. However we do take
into account the crack tip evolution which is reflected in a
cohesive zone size $\epsilon(t_f)$ smaller than $\epsilon_0$. Since
the contact radius is a constant, the crack dwell time $t_r$
(section~\ref{Sec_Cohesive_Zone}) is approximated as the crap tip
age, which is also the contact age. Therefore, when the contact
radius starts decreasing at $t=t_f$, the dwell time $t_r$ is taken
equal to $t_f$.} Inserting Eq. \ref{Eq_SelfCon_Back} into
Eq.~\ref{Eq_GAdhes} provides an approximate criterion for the
begining of the contact radius decrease:
\begin{equation}
g\left(a(t_f),t_f\right) = - \sqrt{\frac{w \pi a_0}{\phi_{1}(t_f)}}
\end{equation}
When the contact radius decreases, Eq.~\ref{deplout} becomes
\begin{equation}
g(a(t),t) = \dot\delta\int_0^{t} d\tau\psi(t-\tau) +
\psi(t)\left(\delta_0 - \frac{a(t)^2}{R} \right) \label{Detg}
\end{equation}
Then $t_f$ is solution to the following implicit equation:
\begin{equation}\label{Eq_Caract}
\psi(t_f)\left(\delta_0-\frac{a_0^2}{R}\right)+\dot{\delta}\int_0^{t_f}d\tau
\psi(t_f-\tau) +\sqrt{\frac{w \pi a_0}{\phi_{1}(t_f)}}=0
\end{equation}

Note that the purely elastic case is recovered for $t=0$. It is also
recovered for constant $\psi(t)$ (elastic material) because
$\delta(t)=\delta_0  + \dot\delta t$, and for long waiting times
where $\psi(t)\simeq\psi(+\infty)$ and
$\int_0^t\psi(\tau)d\tau\simeq t\psi(+\infty)$.

Injecting the initial elastic solution, $t_f$ is identified as the
non zero root of function $f$, with
\begin{equation}
f(t) = \dot{\delta}\int_0^{t}d\tau \psi(t-\tau) + \sqrt{w\pi a_0}
\left(\frac{1}{\sqrt{\phi_{1}(t)}}- \psi(t)\sqrt{\frac{2}{E^*}}
\right) \label{Eq_Simple_Ini}
\end{equation}

Time $t_f$ only depends on the mechanical properties of the material
and on the history of the penetration ($\delta_0$ and
$\dot{\delta}$). Considering a given contact experiment on a rough
surface, the mechanical and adhesive properties as well as the
pull-out velocity are identical for all asperities. Therefore, $t_f$
will only be a function of the initial penetration $\delta_0$ which
is depends on the initial height of a given asperity.

The numerical calculation of $t_f(\delta_0)$ is achieved using a
bisection method. From Eq. \ref{Eq_Simple_Ini}, and knowing that
$a_0$ increases with $\delta_0$, $t_f(\delta_0)$ is a strictly
increasing function of $\delta_0$ and can thus be numerically
inverted to obtain the function $z_f(t)$, representing the initial
penetration necessary for a given asperity to reach the end of the
stick period at time $t$. This function will be useful for the
computation of the total force contribution (section
\ref{Sec_Roughness_Description}) from the asperity height
distribution.

\subsubsection{Decreasing contact radius}\label{Sec_Unloading}
The computation of the load $P_2(t)$ in the case of a decreasing
contact radius is performed using Eq. \ref{Eq_ForceRef} which leads
{(section~\ref{Sec_Force_App_Decrease})} to:
\begin{equation}\label{Eq_ForceDecrease}
P_2(t) = 4a(t)\psi(t) \left(\delta_0-\frac{a(t)^2}{3R}\right) + 4
a(t) \dot{\delta} \int_0^t d\tau \psi(t-\tau)  \label{force2}
\end{equation}

The two terms are respectively the elastic force for a spherical
asperity with the relaxed modulus at time $t$ in presence of
adhesion and the response of the viscoelastic medium to a flat punch
displacement (contact radius $a(t)$).
Note that the purely elastic case is recovered for elastic materials because $\delta(t)=\delta_0+\dot \delta t$.\\
Eqs.~\ref{Eq_SelfCon_Back}, \ref{Eq_dwell}, \ref{Detg} and
\ref{force2} completely describe the decreasing contact radius
phase.

\subsection{Single Asperity -- Normalization} \label{Sec_Normalisation}
For a spherical asperity, following
Maugis~\cite{Barthel02,Maugis92}, the normalization of the contact
variables is performed by introducing the following quantities:
{\begin{eqnarray}\label{}
  \tilde P&=&\frac{P}{\pi w R} \label{Eq_normP},\\
  A&=&\frac{a}{\left(\frac{3 \pi w R^2}{4E^\star}\right)^{1/3}}\label{Eq_normA},\\
  \tilde{\delta}&=&\frac{\delta}{\left(\frac{9 \pi^2 w^2 R}{16 {E^\star}^2}\right)^{1/3}}\label{Eq_normD},\\
  \lambda&=&\frac{2\sigma_0}{\left(\frac{\pi w
  16{E^\star}^2}{9R}\right)^{1/3}}\label{Eq_normL},\\
G(A) &=& -\pst \lambda {\sqrt{2A\tilde{\epsilon}}}.
\end{eqnarray}
}were $\tilde{\epsilon}$ is the normalized size of the cohesive zone
(same length scale as contact radius $a$). In normalized form,
$g(a)$ is given by:
\begin{equation}
G(A_0)= - \frac{\pi}{3} \lambda \sqrt{2A_0\tilde{\epsilon}_0}
\end{equation}

In normalized form, the self-consistent equation
\ref{Eq_SelfCon_Back} is given by:
\begin{equation}
1=\frac {\pi^2}{12}\lambda^2 \tilde{\epsilon}(t) \phit_{1}(tr(t))
\end{equation}
where $\phit_{1}$ is given by:
\begin{equation}
\phit_{1}(t_r)= \frac{2}{{t_r}^2}\int_0^{t_r}d\tau (\tau)\phit(\tau)
\end{equation}

Consequently, Eq. \ref{Eq_Caract} used for the determination of the
end of the first regime now writes:
\begin{equation}
\tilde \psi (t) \left(\tilde \delta(0) - A_0^2\right)+
\dot{\tilde{\delta}}\int_0^{t_f} d\tau\psit(t_f-\tau) + 2
\sqrt{\frac{2A}{3\tilde \phi_{1}(t_f)}}=0
\end{equation}
In the stick period, the force for one isolated asperity (Eq.
\ref{Eq_ForceRef}) is now given by:
\begin{equation}
\tilde P_1(t)= 
    \tilde{\psi}(t) \left(A_0^3(t)-A_0^{1.5}\sqrt6\right) + \frac{3}{2}A_0 \dot{\tilde{\delta}} \int_0^t d\tau\tilde{\psi}(t-\tau) 
\end{equation}
When the contact radius starts decreasing, $G(A(t))$ (Eq.
\ref{Detg}) writes as follow :
\begin{equation}
G(A(t)) = \dot{\tilde{\delta}}\int_0^t d\tau\psit(t-\tau) +
\psit(t)\left(\tilde{\delta}(0) - A^2(t) \right)
\end{equation}
and the force in the second regime is expressed through:
\begin{equation}
\tilde P_2(t) = \tilde{\psi}(t)\left( \frac{3}{2} A(t)\tilde
\delta_0- \frac{A^3(t)}{2}\right) +  \frac{3}{2}A(t)
\dot{\tilde{\delta}} \int_0^t d\tau\tilde{\psi}(t-\tau)
\label{Force2}
\end{equation}

\subsection{Accounting for a rough surface} \label{Sec_Roughness_Description}
We follow the roughness description proposed by Greenwood and
Willamson~\cite{Greenwood66} in their adhesionless elastic contact
study. Namely, a surface of nominal area $A_0$ with $N$ asperities
is in contact with a rigid flat plane. All asperities are assumed to
have the same radius of curvature $R$ and a Gaussian height
distribution of standard deviation $\sigma_s$:
\begin{equation}
\chi(z)=
\frac{1}{\sigma_s\sqrt{2\pi}}\exp\left(-\frac{z^2}{2\sigma_s^2}\right)
\end{equation}
where $\chi(z)$ is the probability that the summit of an asperity
stands between z and $z+dz$. {For a penetration of the rigid flat
$\delta_0$ }, the number of asperities in contact is given by:
$n=N\int_{\delta_0}^\infty dz\chi(z)$ and all contact variables can
be obtained with the same integration. Similarly, the calculation of
the load at time $t$ is performed by integrating the viscoelastic
response of each asperity initially in contact: {\begin{equation}
P(t) =N\int_{0}^\infty dz\chi(z+\delta_0)P_a(z,t)  \label{IntTot}
\end{equation}
}where $P_a(z,t)$ is the load at time $t$ of an asperity undergoing
an initial penetration of {$z-\delta_0$}.

\section{Results and discussion} \label{Sec_Results}

\subsection{Single asperity -- Impact of pull-out velocity}  \label{Sec_Single_Asperiy}
Force {\it vs} penetration curves (force plots) for a single
asperity were calculated with $\lambda$=5 { -- a value consistent
with our assumption of small but finite cohesive zone size -- } and
for a moderately viscoelastic material with $k=0.5$.
Figures~\ref{Fig_FD_Vitesse_Lent} and \ref{Fig_FD_Vitesse_Rapide}
display results obtained for $\tilde\delta_0=5$ for pull-out
velocities $\tilde{\dot \delta}$ ranging from $-0.01$ to $-50$.

For very low pull-out velocity values ($|\tilde{\dot \delta}| \leq
0.1$), there is an initial rapid decrease of the force, followed by
a linear variation of the force as a function of penetration. This
is due to the fast relaxation of the (predominantly compressive)
stresses. The stress field is then controlled by the relaxed modulus
which leads to constant contact stiffness and a linear
force-distance curve segment, before the contact radius starts
decreasing. {When the penetration decreases further}, the force
becomes tensile (negative), displays an incurvation, levels off,
finally increases and jumps to zero: this evolution of the force
results from contact edge recession and asperity snap-off. The
absolute value of the maximum tensile force defines the pull-out
force $F_{pull out}$, and the (positive) effective adhesion
$w_{eff}$ through {$ w_{eff} = F_{pull out}/ (3/2\ \pi R)$}.

The very low velocity case is an elastic adhesive contact with an
effective modulus equal to the relaxed modulus and a pull-out force
equal to 3/2 which is the JKR value. At slightly larger pull-out
velocities, we observe that the pull out force {\em increases} with
pull-out velocity from this 3/2 value: higher pull-out velocity
leads to increased contact radius velocity. Enhancement of the
viscoelastic crack tip dissipation and effective adhesion follow as
amply demonstrated in the
literature\cite{Schapery75,Schapery75bis,Greenwood81,Hui98,Lin99,Schapery89,Greenwood06}.
It results from the combination of a fully relaxed contact zone and
a less relaxed cohesive zone.

In contrast, for large pull-out velocities ($|\tilde{\dot \delta}|
\geq 1$), the initial precipitous drop of the force is no longer
present and the pull-out force {\em decreases} with pull-out
velocity. Eventually for very large pull-out velocity ({\it e.g.}
$\tilde{\dot \delta}=-50$) the behavior tends toward an elastic
behavior and the pull out force reverts to 3/2. Indeed, for large
pull-out velocities, the relaxation of the compressive stresses in
the contact zone is far from complete during the experimental time.
The pull-out force is lower than for low pull-out velocity because
the compressive contact stresses favor contact rupture.

As a consequence, there exists a transition with an optimum pull-out
velocity (here of around $\tilde{\dot \delta}=-1$) for which the
pull-out force is maximum. This result specifically originates from
the consistent inclusion of both viscoelastic crack tip dissipation
and contact stress relaxation.

\subsection{Rough surfaces}
As an {example} of  results for rough surfaces, the same material
parameters $\lambda=5$ and $k=0.5$ are chosen. The initial
penetration is $\tilde{\delta_0}=5$.

\subsubsection{Force distribution}
{Normalizing the roughness as the penetration (Eq.~\ref{Eq_normD}),
we compute the force curve for a rough surface with a normalized
standard deviation $\tilde{\sigma_s}=2$.} Computing the
contributions of asperities displaying each regime (stick period
\emph{vs.} decreasing contact radius) to the total force for
pull-out velocity $\tilde{\dot \delta}=-1$ (Figure
\ref{CompletRef}), we observe that the proportion of asperities
initially in contact is almost $100 \%$, as expected for such a
ratio of initial penetration $\tilde{\delta_0}$ to roughness
$\tilde{\sigma_s}$. {This type of contact is relevant to glass
molding where a rather homogeneous distribution of asperities, with
narrow asperity height and radius distributions, is generated on the
mold surface. Note that the Greenwood-Williamson model is
particularly apt at treating such a surface morphology. It would be
less adequate for a less regular surface with a wide distribution of
asperity height and/or a wide distribution of asperity radius.}

The force is initially dominated by asperities pinned in the stick
period (black dashed line). A gradual transition where the contact
zones of the lower asperities {de-pin} and start to recede is
observed: the contribution of pinned asperities drops to zero. Then
the lowest asperities start to undergo contact rupture. It is in the
middle of this transition that the maximum tensile force (pull-out
force) is recorded.

\subsubsection{Impact of roughness -- Constant pull-out velocity}

Force plots with increasing roughness (from $\tilde{\sigma_s}=0.1$
to $\tilde{\sigma_s}=5$) are displayed on Figure
\ref{Fig_VariationET}. For small $\tilde{\sigma_s}$, the results
track the single asperity limit: all the asperities in the
population exhibit an identical behaviour, having nearly the same
initial penetration. For larger $\tilde{\sigma_s}$, the initial
contact force increases as expected from the elastic (JKR) limit:
for a single asperity the increase of the force with initial
penetration is more rapid than linear. Simultaneously, the duration
of the contact increases and the pull-out force decreases with
roughness as in the elastic case~\cite{Fuller75}. Indeed, for one
isolated asperity, the pull-out time increases with initial
penetration. Therefore, when $\tilde{\sigma_s}$ increases, the
distribution of the pull-out times spreads out and the different
asperities contribute less constructively to the total pull-out
force, which therefore decreases.

\subsubsection{Impact of pull-out velocity -- Constant roughness}

Figures~\ref{Fig_VariationVb} and \ref{Fig_VariationVa} display
force plots at different pull-out velocity values, from $\tilde{\dot
\delta}=-0.01$ to $\tilde{\dot \delta}=-10$ {for a normalized
standard deviation $\tilde{\sigma_s}=2$.} The overall picture is
similar to the single asperity case
(section~\ref{Sec_Single_Asperiy}) and the same transition from low
to high velocity regimes is observed. At low pull-out velocities, a
fast decrease of the force due to stress relaxation is followed by a
quasi elastic behavior while at large velocities, the system is
purely elastic. The pull-out forces in these two limit regimes are
differentiated by the roughness. This is because the elastic moduli
involved are different: indeed, the {\em relaxed} modulus is
pertinent for the very low pull-out velocity regime, and the {\em
instantaneous} modulus for the very high velocity regime. A larger
elastic modulus will result in a larger adhesion parameter and a
lower adhesion force~\cite{Fuller75,Maugis96}. A pull-out force
enhancement is recorded between these two limit regimes because of
the combination of partial contact stress relaxation and crack tip
dissipation.

It is interesting to assess the details of the transition between
these two regimes. Calculation of pull-out forces as a function of
pull-out velocities were performed for a rough surface with a
roughness of $\tilde{\sigma_s}=2$, a cohesive zone parameter
$\lambda=5$ and a moderately viscoelastic material $k=0.5$. The
pull-out force as a function of pull-out velocity has been
normalized to the rough surface limit in the high velocity elastic
regime (instantaneous modulus). The results are displayed on
Fig.~\ref{Fig_Enhancement} where the isolated asperity result is
also shown for comparison. As expected both exhibit enhancement of
the pull-out force in the intermediate regime where contact stress
relaxation and crack tip creep occur at the same time.
Figure~\ref{Fig_Enhancement} clearly evidences that the enhancement
of adhesion is much more pronounced on rough viscoelastic surfaces
than for an isolated asperity. Indeed for elastic materials a
roughness as large as $\tilde{\sigma_s}=2$ severely depresses the
pull-out force due to the stored elastic energy of the higher
asperities. However for a viscoelastic material these elastic
stresses rapidly relax while the single asperity adhesive force is
enhanced through crack tip dissipation. As a result, even a very
moderately viscoelastic material is sufficient to enhance the
effective adhesion on rough surfaces by an order of magnitude over a
large range of pull out forces (Figure~\ref{Fig_Enhancement}).

\section{Conclusion}
A simple model for the adhesive contact of rough viscoelastic
surfaces has been developed based on a Greenwood-Williamson
roughness distribution. For that purpose a simplified model has been
proposed for the adhesive contact of viscoelastic asperities,
assuming instantaneous loading and constant pull-out velocity. It
retains the full phenomenology of the viscoelastic contact including
stress relaxation inside the contact zone and creep in the cohesive
zone.

For elastic surfaces, it is well known that increasing roughness
rapidly suppresses adhesion. Our calculations give a quantitative
assessment of how efficiently viscoelasticity restores adhesion on
rough surfaces through the interplay between contact stress
relaxation and cohesive zone creep. The phenomenon operates in a
wide range of pull-out velocities.

\section{Acknowledgements}
We thank I. Sharma for numerous suggestions.

{
\section{Appendix -- Computation of the total force}\label{Sec_Force_App}
The total force writes as \cite{Haiat03}:
\begin{equation}\label{Eq_ForceGeneral}
P(t)=4 \int_0^{+\infty} g(r,t) dr
\end{equation}
The contribution of the cohesive zone is of higher order in
$\epsilon$ (cohesive zone size) and is neglected, which is typical
for large $\lambda$. Then Eq.~\ref{Eq_ForceGeneral} becomes $P(t)=4
\int_0^{a(t)} g(r,t) dr$ and Eq.~\ref{gdth} becomes
\begin{equation}
g(r,t)=\int_{t_{a-}(a(t))}^t d\tau \psi(t-\tau) \dsdd{\tau}
\theta(r,\tau) \label{Eq_gdth_simpl}
\end{equation} with $\theta(r,\tau)=\delta(\tau)-\delta_0(r)$ which is
Eq.~\ref{defth2}. Inserting Eq.~\ref{Eq_gdth_simpl}
into~\ref{Eq_ForceGeneral} and inverting the order of integration
(cf also Ref.~\cite{Haiat03}, section 3.2.4) results in
Eq.~\ref{Eq_ForceRef}.
\subsection{During the stick period}\label{Sec_Force_App_Stick}
During the stick period by assumption $a(t)=a_0$ is constant and
$\min(a(t),a(\tau))=a_0$. In addition for a sphere of radius $R$
loaded as specified in section~\ref{Sec_Boundary_Conditions},
combining Eqs.~\ref{defth2} and \ref{Eq_Depl_Unload}
\begin{equation}
\theta(r,\tau)=(\delta_0+\tau\dot\delta -r^2/R)\Theta(\tau)
\end{equation}
Then Eq.~\ref{Eq_ForceRef} gives:
\begin{eqnarray}
P_1(t) &=&  4 \psi(t) \int_0^{a_0} dr (\delta_0-\frac{r^2}{R})  +  4 a_0 \dot{\delta} \int_0^t d\tau \psi(t-\tau) \\
P_1(t) &=&  4 \psi(t) (a_0 \delta_0-\frac{a_0^3}{3R})  +  4 a_0
\dot{\delta} \int_0^t d\tau \psi(t-\tau) \end{eqnarray} where $P_1$
denotes the force during the stick period. The initial penetration
$\delta_0$ is given by the double Hertz model and writes:
\begin{equation}
\delta_0=\frac{a_0^2}{R}- \frac{\sigma_0 \pi \sqrt{a_0
\epsilon_0}}{\sqrt{2}E^*}.
\end{equation}
This last equation, in combination with the previous one, gives
Eq.~\ref{Eq_ForceStick}.
\subsection{Decreasing contact radius}\label{Sec_Force_App_Decrease}
By assumption the contact radius is constant or monotonously
decreasing from $\tau=0$ so that $\min(a(t),a(\tau))=a(t)$ and
Eq.~\ref{Eq_ForceRef} directly results in
Eq.~\ref{Eq_ForceDecrease}.}

\newpage

\providecommand{\refin}[1]{\\ \textbf{Referenced in:} #1}

\section*{Captions and Figures}
\begin{figure}[!ht]
\includegraphics[width=10cm]{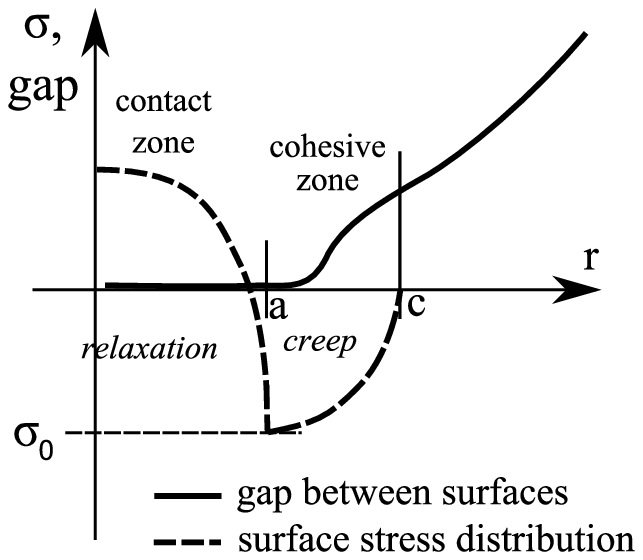}\caption{}\label{Fig_Contact_Model}
\end{figure}

Fig.~\ref{Fig_Contact_Model}: Schematics of the viscoelastic
adhesive contact showing the gap between the surfaces and the normal
surface stress distribution. Special emphasis is given to the
cohesive zone where the attractive interactions operate across the
gap between surfaces.

\begin{figure}[!ht]
\includegraphics[width=10cm]{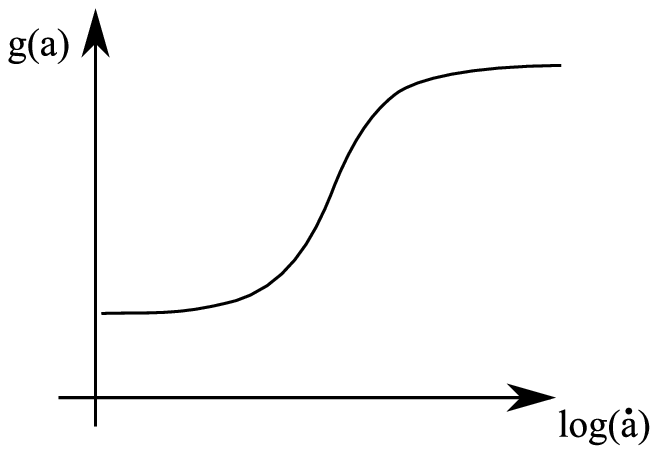}\caption{}\label{Fig_Outer_K}
\end{figure}

Fig.~\ref{Fig_Outer_K}: Schematics of the relation between $g(a)$ --
proportional to the stress intensity factor -- and the crack
velocity (after~\cite{Barthel02}).

\begin{figure}[!ht]
\includegraphics[width=10cm]{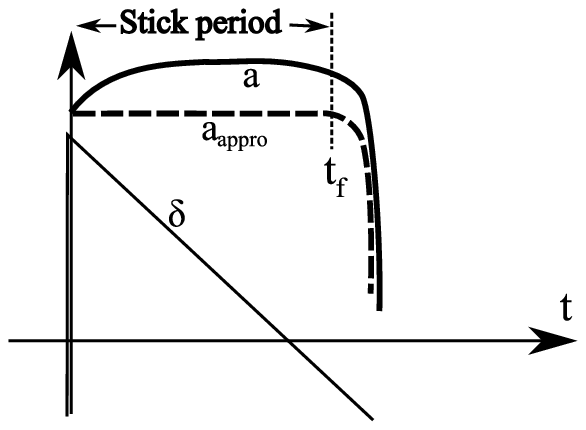}
\caption{} \label{Fig_Histoire_Pene}
\end{figure}

Fig.~\ref{Fig_Histoire_Pene}: Prescribed penetration as a function
of time ($\delta(t)$) and resulting contact radius history ($a(t)$).
The constant radius approximation proposed in the present model is
also shown as $a_{appro}(t)$. The end of the stick period is $t_f$.

\begin{figure}[!ht]
\includegraphics[width=12cm]{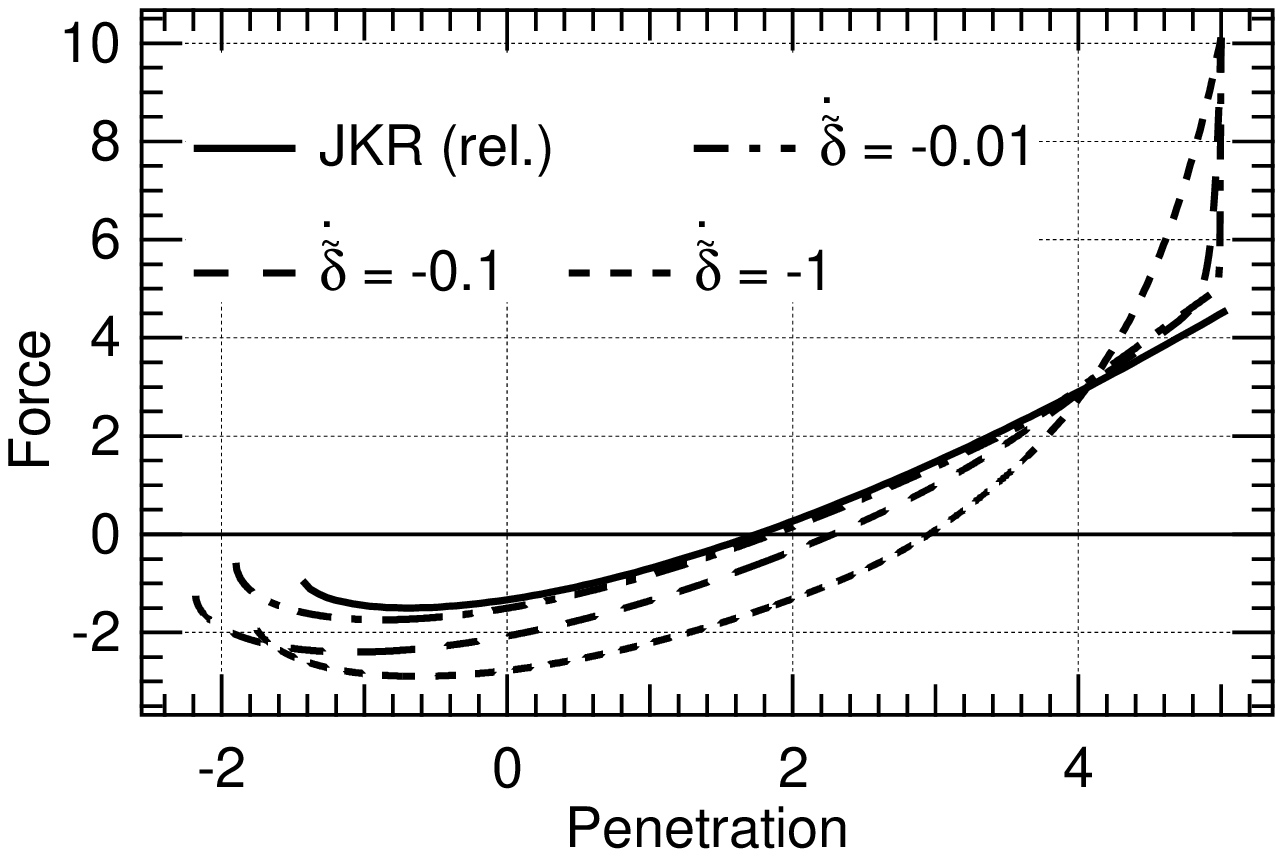} \caption{}
\label{Fig_FD_Vitesse_Lent}
\end{figure}

Fig.~\ref{Fig_FD_Vitesse_Lent}: Single asperity force plot for
different values of the pull-out velocity (from $\tilde{\dot
\delta}=-0.01$ to $\tilde{\dot \delta}=-1$) and for the elastic
limit (JKR) with the relaxed modulus.

\begin{figure}[!ht]
\includegraphics[width=12cm]{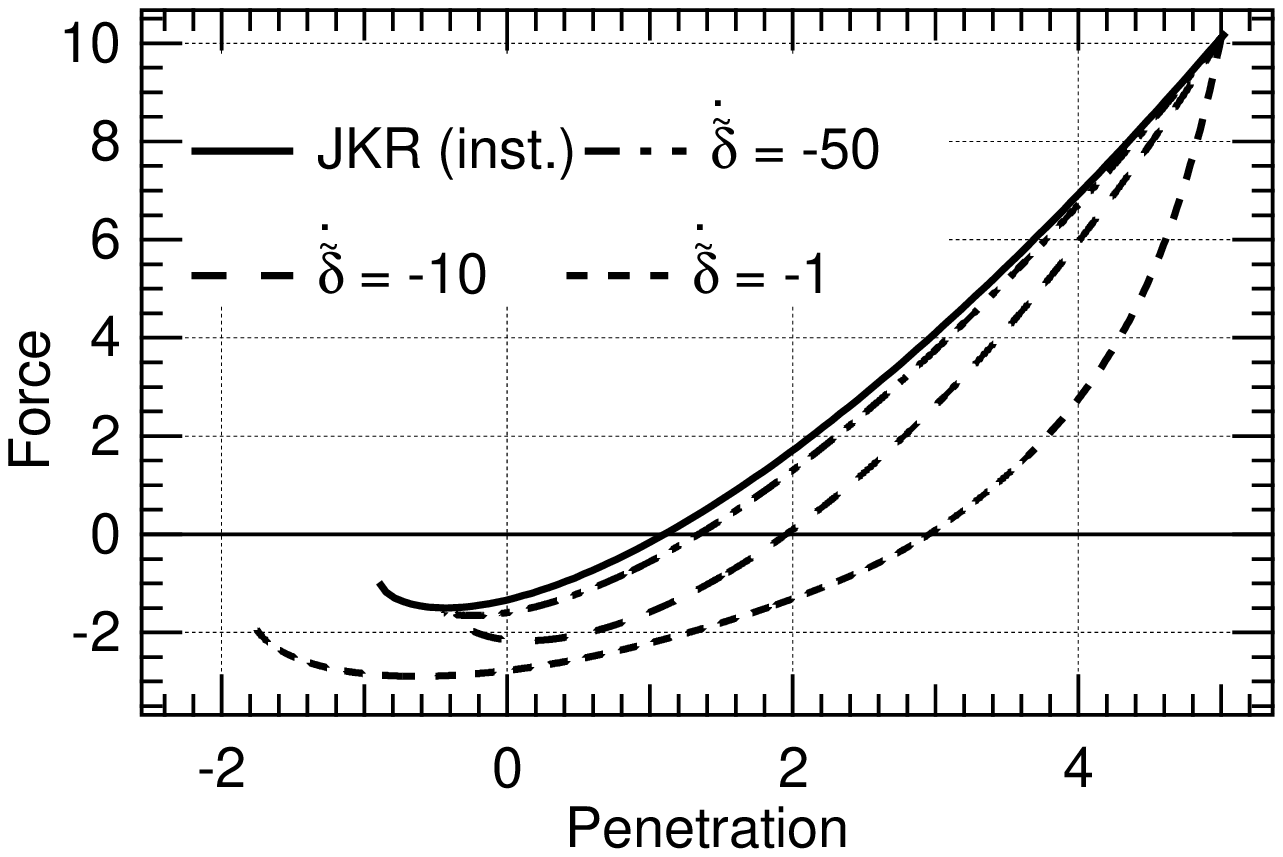} \caption{}
\label{Fig_FD_Vitesse_Rapide}
\end{figure}

Fig.~\ref{Fig_FD_Vitesse_Rapide}: Single asperity force plot for
different values of the pull-out velocity (from $\tilde{\dot
\delta}=-1$ to $\tilde{\dot \delta}=-50$) and for the elastic limit
(JKR) with the instantaneous modulus.

\begin{figure}[!ht]
\includegraphics[width=12cm]{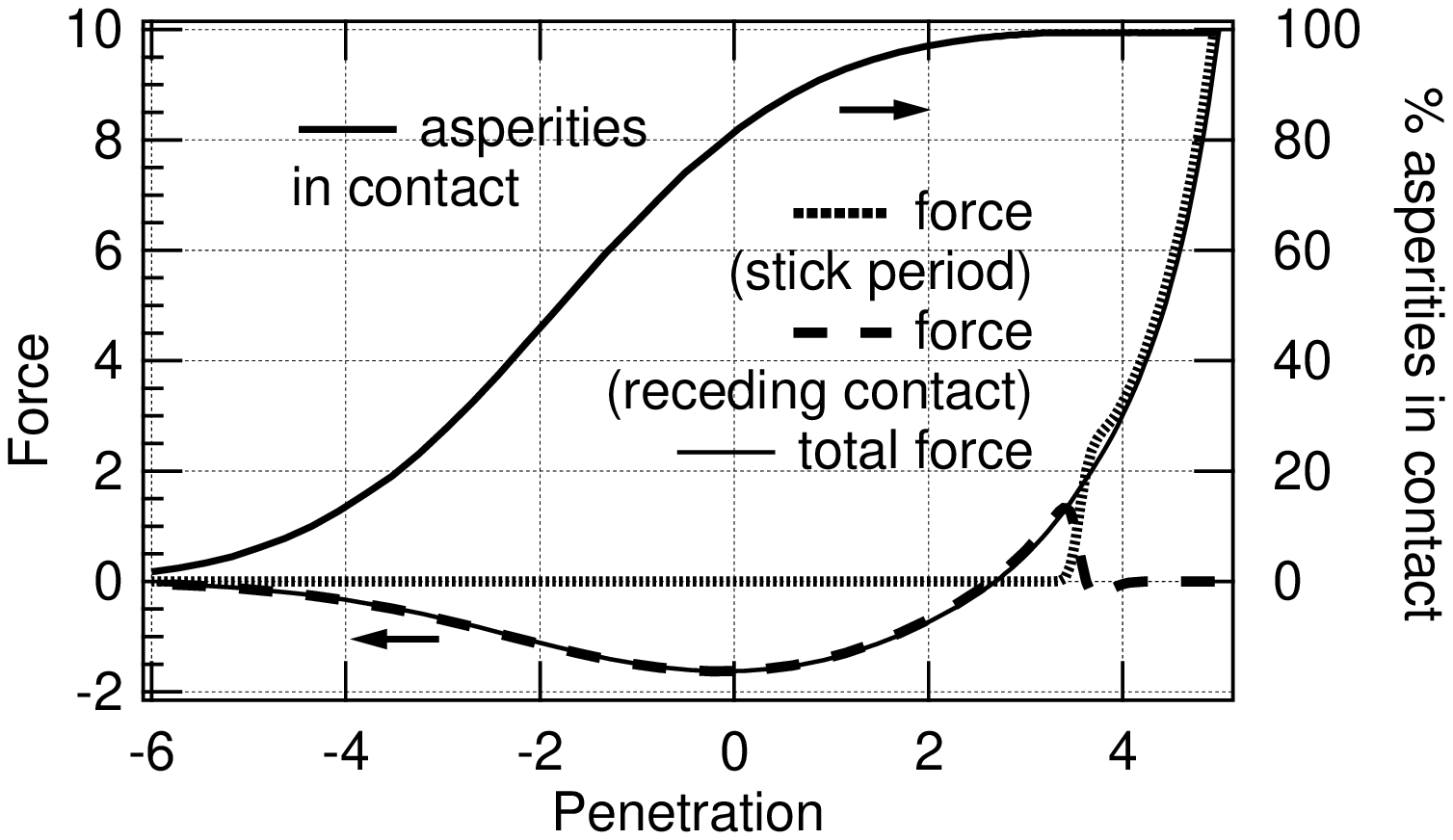} \caption{}
\label{CompletRef}
\end{figure}

Fig.~\ref{CompletRef}: Force distribution during adhesion rupture on
a rough surface showing the contributions from asperities in the
stick period and asperities with receding contact radius. The
proportion of asperities in contact is shown on the right hand axis.

\begin{figure}[!ht]
\includegraphics[width=12cm]{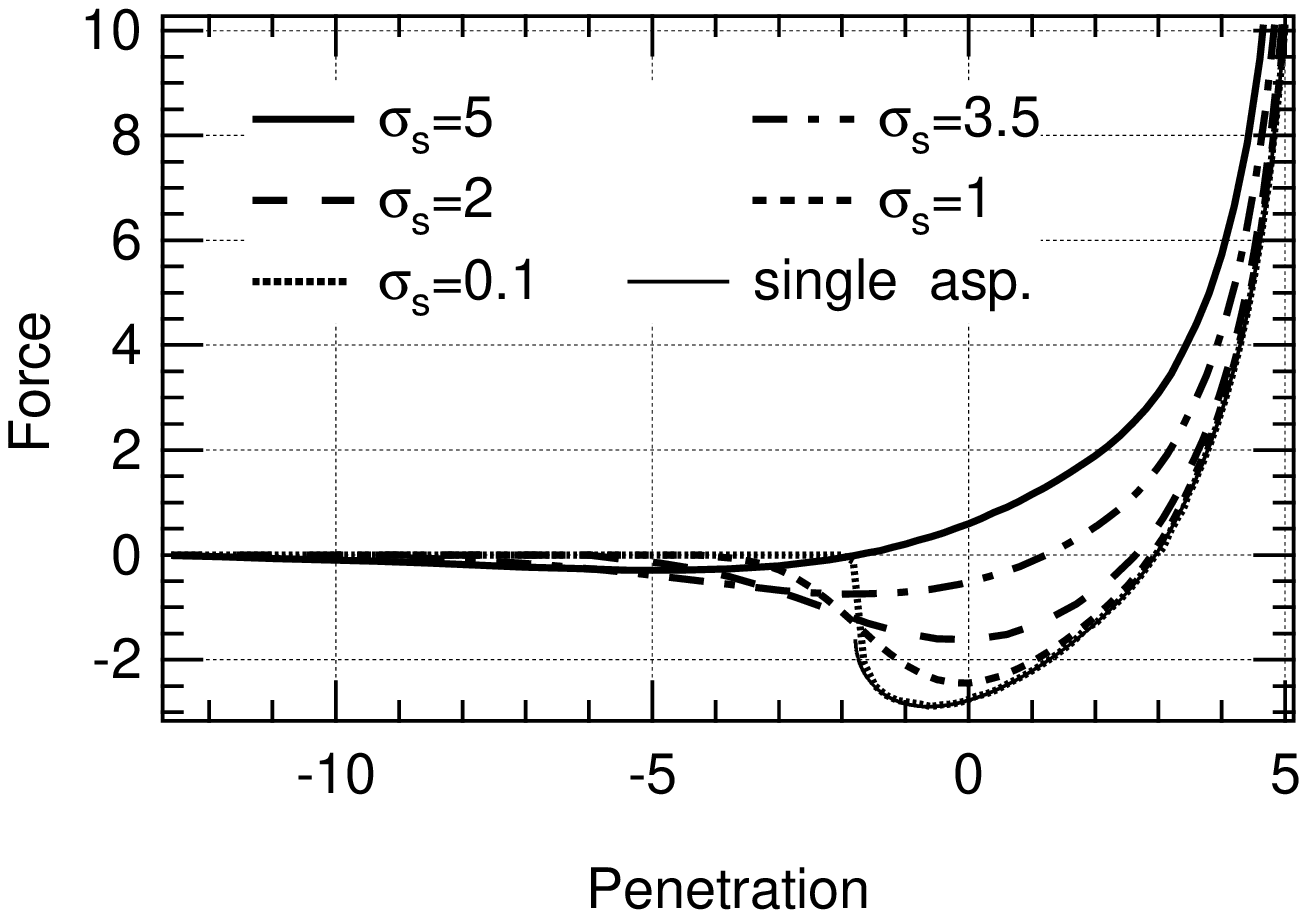}
\caption{} \label{Fig_VariationET}
\end{figure}

Fig.~\ref{Fig_VariationET}: Force plots for rough surfaces with
increasing roughness. Here $\delta_0=5$ and $\dot \delta = -1$.

\begin{figure}[!ht]
\includegraphics[width=12cm]{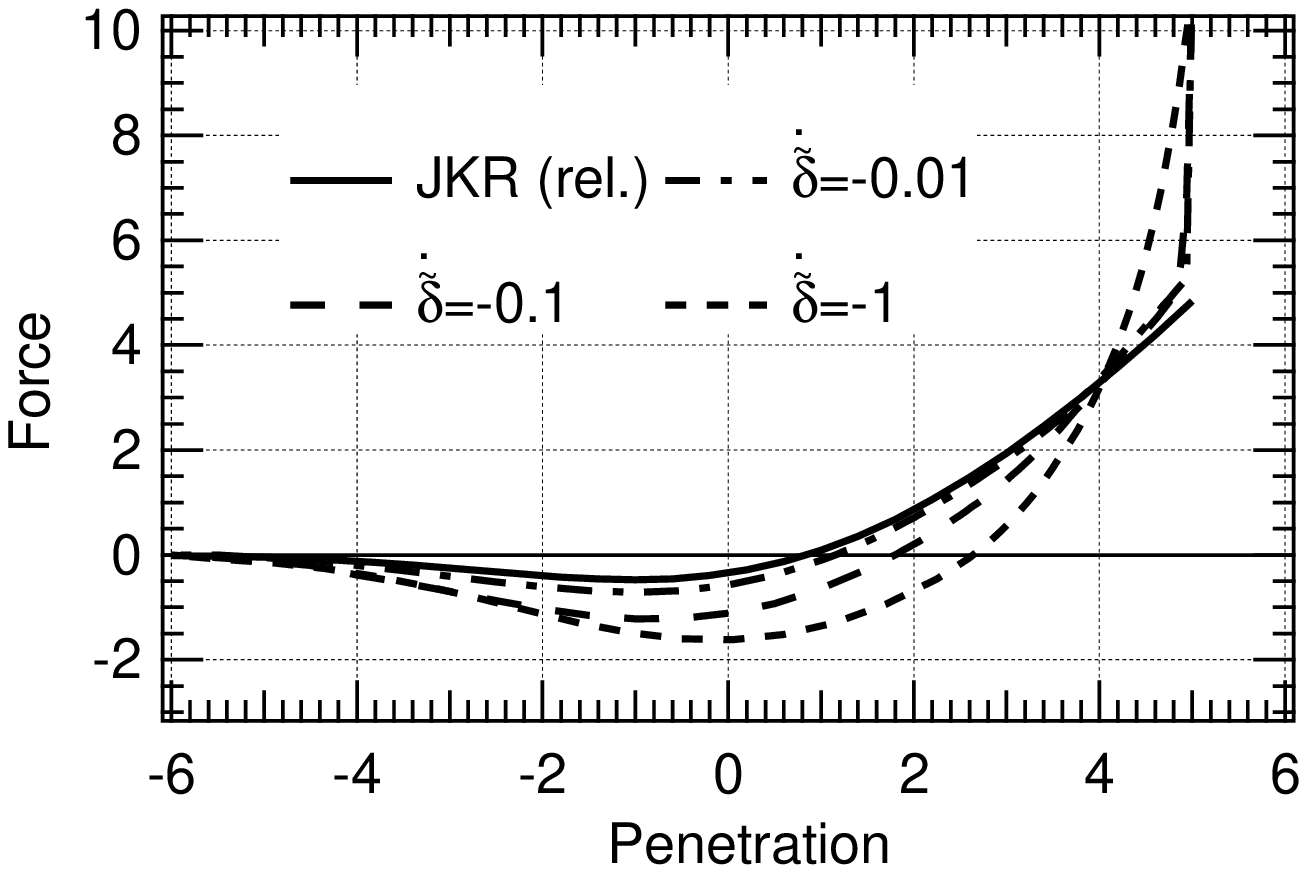}
\caption{} \label{Fig_VariationVb}
\end{figure}

Fig.~\ref{Fig_VariationVb}: Force plots for a rough surface with
$\sigma_s=2$ for increasing pull-out velocity at low velocities.
Also shown is the elastic limit for the relaxed modulus.

\begin{figure}[!ht]
\includegraphics[width=12cm]{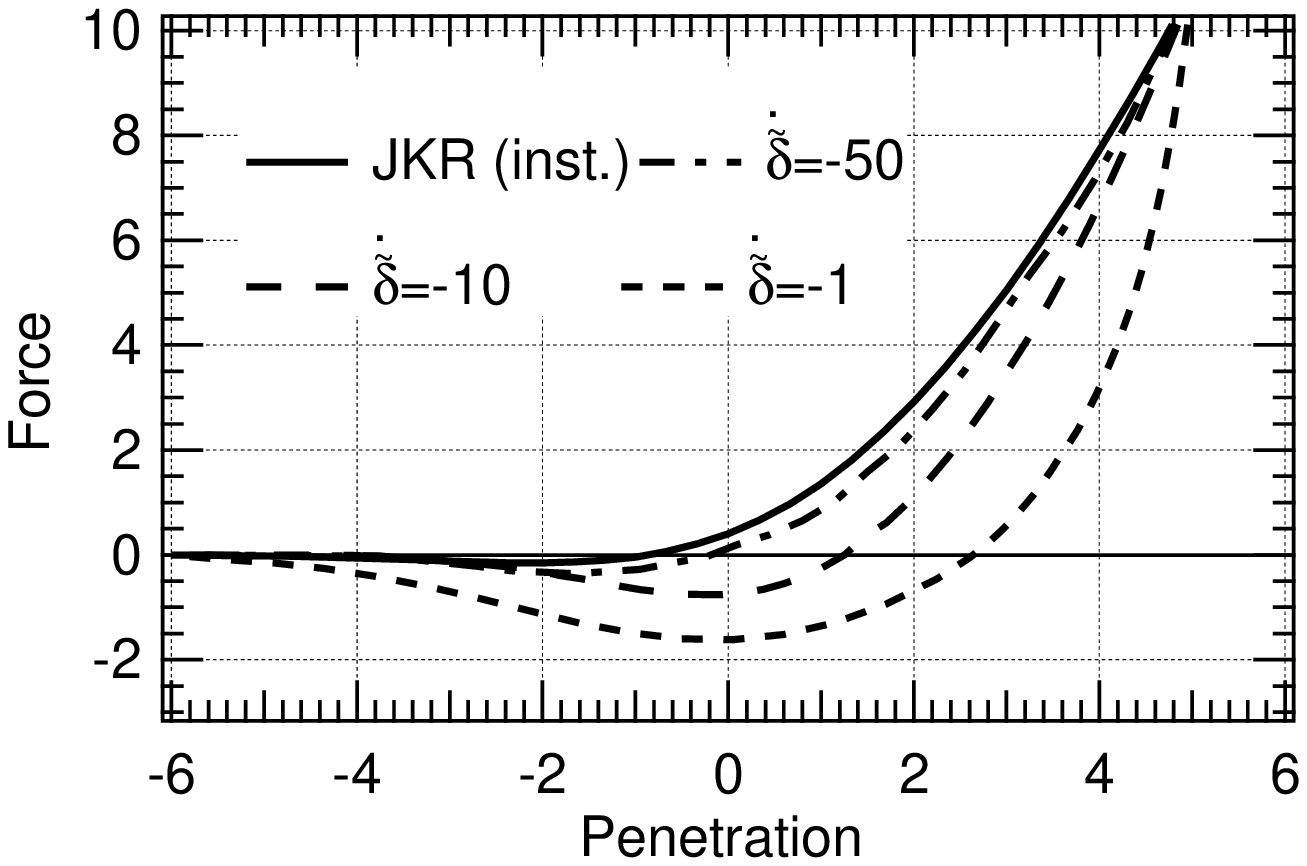}
\caption{} \label{Fig_VariationVa}
\end{figure}

Fig.~\ref{Fig_VariationVa}:  Force plots for a rough surface with
$\sigma_s=2$ for increasing pull-out velocity at high velocities.
Also shown is the elastic limit for the instantaneous modulus.

\begin{figure}[!ht]
\includegraphics[width=12cm]{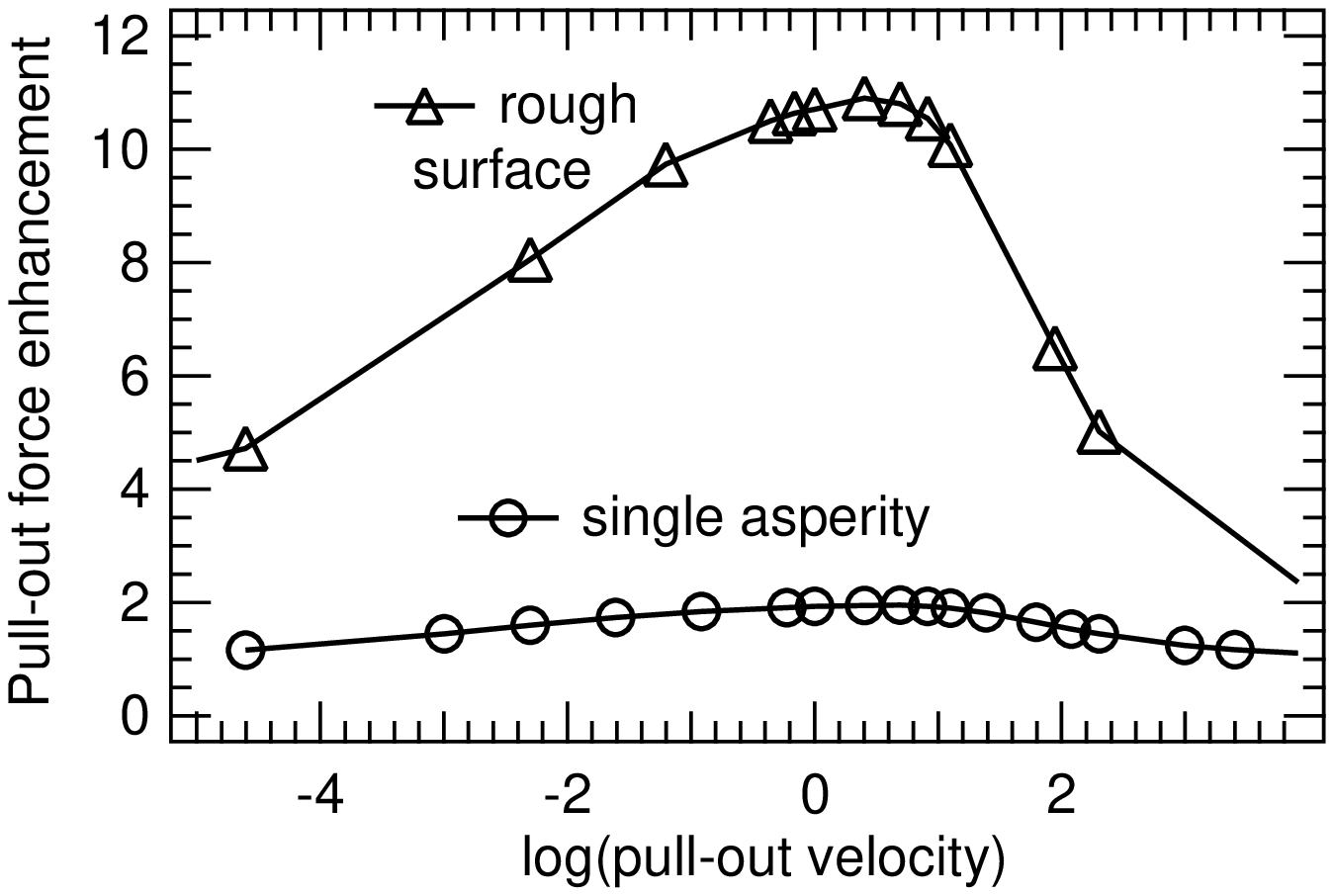}
\caption{} \label{Fig_Enhancement}
\end{figure}

Fig.~\ref{Fig_Enhancement}: Pull-out force enhancement over elastic
case as a function of pull-out velocity for an isolated asperity and
for a rough surface with $\sigma_s=2$.


\begin{center}
 \begin{figure}[!ht]
\includegraphics[width=8.9cm]{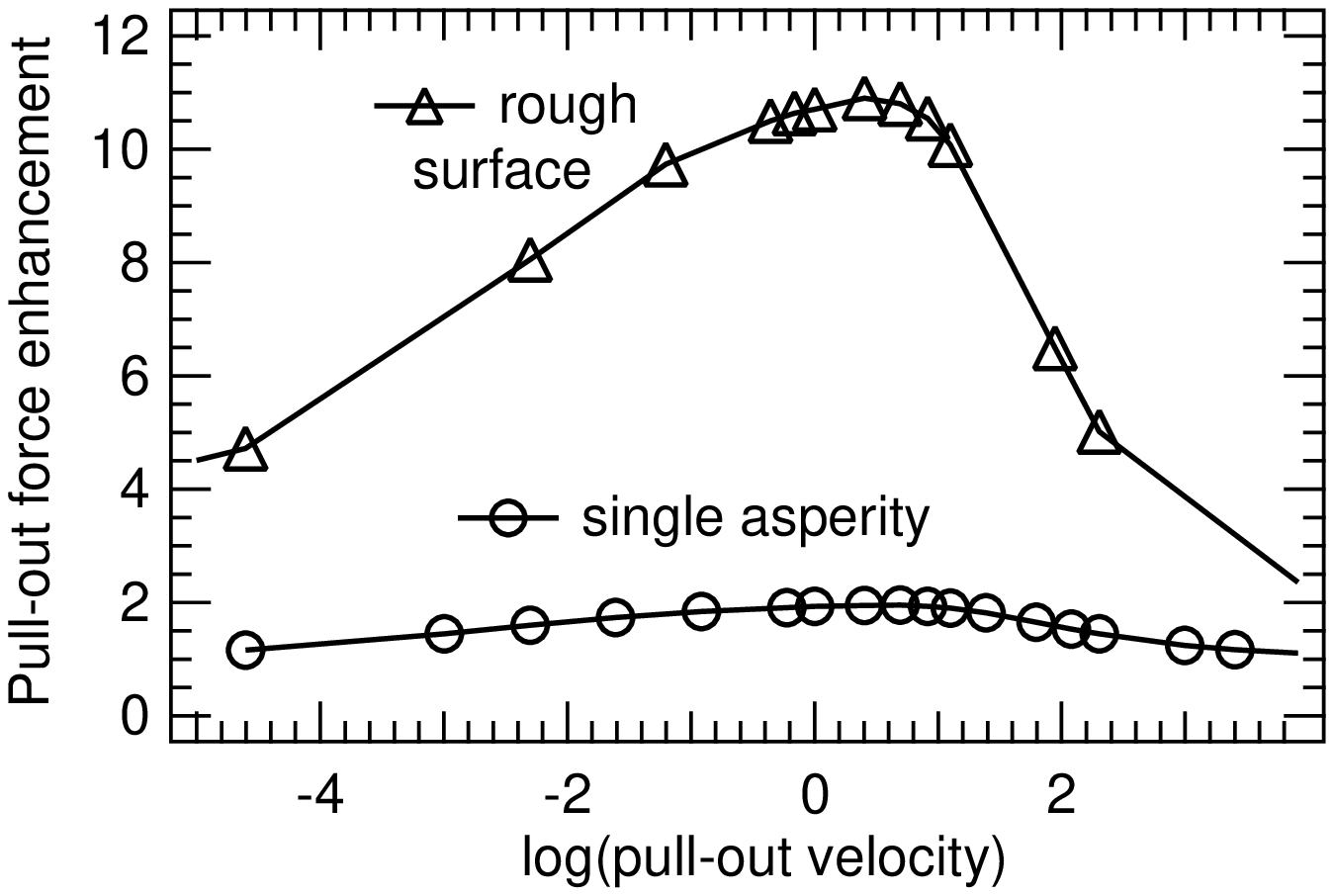}
\caption{Table of Contents Graphic}
\end{figure}
\end{center}


\begin{thebibliography}{10}

\bibitem{JUANG01}
Juang~Y.~J.; Bruer~D.; Lee~L.~J.; Koelling~K.~W.; Srinivasan~N.;
Drummond~C.~H.; Wong~B.~C. \emph{J. Appl. Polym. Sci} \textbf{2001},
\emph{80}, 521.

\bibitem{ZOSEL97}
Zosel,~A. \emph{J. Adhesion Sci. Technol.} \textbf{1997}, \emph{11},
1447.

\bibitem{GAY99}
Gay,~C.; Leibler~L. \emph{Phys. Rev. Lett.} \textbf{1999},
\emph{82}, 936.

\bibitem{PECH05}
Pech~J.; Berthome~G.; Jeymond~M.; Eustathopoulos~N. \emph{Glass Sci.
Technol.} \textbf{2005}, \emph{78}, 54.

\bibitem{Fuller75}
Fuller,~K.; Tabor,~D. \emph{Proc. Roy. Soc. A} \textbf{1975},
\emph{345}, 327.

\bibitem{Ting66}
Ting,~T. C.~T. \emph{J. Appl. Mech} \textbf{1966}, \emph{33}, 845.

\bibitem{Schapery75}
Schapery,~R.~A. \emph{Int J Fract} \textbf{1975}, \emph{3}, 369.

\bibitem{Schapery75bis}
Schapery,~R.~A. \emph{Int J Fract} \textbf{1975}, \emph{11}, 549.

\bibitem{Greenwood81}
Greenwood,~J.~A.; Johnson,~K.~L. \emph{Phil. Mag.} \textbf{1981},
\emph{43},
  697.

\bibitem{Greenwood04}
Greenwood,~J.~A. \emph{J. phys. D, Appl. phys.} \textbf{2004},
\emph{37},
  2557.

\bibitem{Hui98}
Hui,~C.~Y.; Baney,~J.~M.; Kramer,~E.~J. \emph{Langmuir}
\textbf{1998},
  \emph{14}, 6570.

\bibitem{Lin99}
Lin,~Y.~Y.; Hui,~C.~Y.; Baney,~J.~M. \emph{J. phys., D, Appl. phys.}
  \textbf{1999}, \emph{32}, 2250.

\bibitem{Hui00}
Hui,~C.~Y.; Lin,~Y.~Y.; Baney,~J.~M. \emph{J. polym. sci., Part B,
Polym. phys}
  \textbf{2000}, \emph{38}, 1485.

\bibitem{Barthel02}
Barthel,~E.; Haiat,~G. \emph{Langmuir} \textbf{2002}, \emph{18},
9362.

\bibitem{Haiat03}
Haiat,~G.; Phan~Huy,~M.~C.; Barthel,~E. \emph{J. Mech. Phys. Sol.}
  \textbf{2003}, \emph{51}, 69.

\bibitem{Barthel04}
Barthel,~E.; Haiat,~G. \emph{J. Adhesion} \textbf{2004}, \emph{80},
1.

\bibitem{Schapery89}
Schapery,~R.~A. \emph{Int. J. Fract.} \textbf{1989}, \emph{39}, 163.



\bibitem{LAKES06}
Lakes, R. S. and Wineman A.  \emph{J. Elasticity} \textbf{2006},
\emph{85}, 45.

\bibitem{Barthel07}
Barthel,~E.; Perriot,~A. \emph{J. Phys. D} \textbf{2007}, \emph{40},
  1059.

\bibitem{Barthel98}
Barthel,~E. \emph{J. Colloid Interface Sci.} \textbf{1998},
\emph{200}, 7.

\bibitem{Greenwood98}
Greenwood,~J.~A.; Johnson,~K.~L. \emph{J. Phys. D: Appl. Phys.}
\textbf{1998},
  \emph{31}, 3279.

\bibitem{Maugis92}
Maugis,~D. \emph{J. Colloid Interface Sci} \textbf{1992},
\emph{150}, 243.

\bibitem{Greenwood66}
Greenwood,~J.; Williamson,~J. \emph{Proc. Roy. Soc. A}
\textbf{1966},
  \emph{295}, 300.

\bibitem{Greenwood06}
Greenwood,~J.~A.; Johnson,~K.~L. \emph{J. Colloid Interface Sci}
\textbf{2006},
  \emph{296}, 284.

\bibitem{Maugis96}
Maugis,~D. \emph{J. Adhes. Sci. Technol.} \textbf{1996}, \emph{10},
161.

\end{thebibliography}
\end{document}